\documentclass[onecolumn]{aastex}

\published{\today~- Version 1.0}
\begin{document}

\title{Large Synoptic Survey Telescope Solar System Science Roadmap}


\author{Megan E. Schwamb}
\affiliation{Gemini Observatory,  Northern Operations Center, 670 North A'ohoku Place, Hilo, HI 96720, USA}

\author{R. Lynne Jones}
\affiliation{Department of Astronomy, University of Washington, 3910 15th Ave NE, Seattle, WA 98195, USA}

\author{Steven R. Chesley}
\affiliation{Jet Propulsion Laboratory, California Institute of Technology, 4800 Oak Grove Drive, Pasadena, CA, 91109, USA}

\author{Alan Fitzsimmons}
\affiliation{Astrophysics Research Centre, School of Mathematics and Physics, Queen's University Belfast, Belfast BT7 1NN, UK}

\author{Wesley C. Fraser}
\affiliation{Astrophysics Research Centre, School of Mathematics and Physics, Queen's University Belfast, Belfast BT7 1NN, UK}

\author{Matthew J. Holman}
\affiliation{Harvard-Smithsonian Center for Astrophysics, 60 Garden St., MS 51, Cambridge, MA 02138, USA}

\author{Henry Hsieh}
\affiliation{Planetary Science Institute, 1700 East Fort Lowell Road, Suite 106, Tucson, AZ 85719, USA}

\author{Darin Ragozzine}
\affiliation{Brigham Young University, Department of Physics and Astronomy, N283 ESC, Provo, UT 84602, USA}

\author{Cristina A. Thomas}
\affiliation{Planetary Science Institute, 1700 East Fort Lowell Road, Suite 106, Tucson, AZ 85719, USA}
\affiliation{Department of Physics and Astronomy, Northern Arizona University, P.O. Box 6010, Flagstaff, AZ 86011, USA}

\author{David E. Trilling}
\affiliation{Department of Physics and Astronomy, Northern Arizona University, P.O. Box 6010, Flagstaff, AZ 86011, USA}

\author{Michael E. Brown}
\affiliation{Division of Geological and Planetary Sciences, California Institute of Technology, Pasadena, CA 91125, USA}

\author{Michele T. Bannister}
\affiliation{Astrophysics Research Centre, School of Mathematics and Physics, Queen's University Belfast, Belfast BT7 1NN, UK}

\author{Dennis Bodewits}
\affiliation{Department of Astronomy, University of Maryland, College Park, MD 20742-2421, USA}

\author{Miguel de Val-Borro}
\affiliation{NASA Goddard Space Flight Center, Astrochemistry Laboratory, 8800 Greenbelt Road, Greenbelt, MD 20771, USA}
\affiliation{Department of Physics, Catholic University of America, Washington, DC 20064, USA}

\author{David Gerdes}
\affiliation{ Department of Physics, University of Michigan, Ann Arbor, MI 48109, USA}

\author{Mikael Granvik}
\affiliation{Department of Physics, University of Helsinki, P.O. Box 64, 00014 University of Helsinki, Finland}

\author{Michael S. P. Kelley}
\affiliation{Department of Astronomy, University of Maryland, College Park, MD 20742-2421, USA}

\author{Matthew M. Knight}
\affiliation{Department of Astronomy, University of Maryland, College Park, MD 20742-2421, USA}

\author{Robert L. Seaman}
\affiliation{Lunar and Planetary Laboratory, University of Arizona, 1629 E. University Blvd., Tucson, AZ 85721, USA}

\author{Quan-Zhi Ye}
\affiliation{Division of Physics, Mathematics and Astronomy, California Institute of Technology, Pasadena, CA 91125, USA}
\affiliation{Infrared Processing and Analysis Center, California Institute of Technology, Pasadena, CA 91125, USA}

\author{Leslie A. Young}
\affiliation{Southwest Research Institute, 1050 Walnut St., Suite 300, Boulder CO, 80302, USA}

 \collaboration{on behalf of the LSST Solar System Science Collaboration}

\correspondingauthor{Megan E. Schwamb}
\email{mschwamb.astro@gmail.com}


\begin{abstract}

The Large Synoptic Survey Telescope (LSST) is uniquely equipped to search for Solar System bodies due to its unprecedented combination of depth and wide field coverage. Over a ten-year period starting in 2022, LSST will generate the largest catalog of Solar System objects to date. The main goal of the LSST Solar System Science Collaboration (SSSC) is to facilitate the efforts of the planetary community to study the planets and small body populations residing within our Solar System using LSST data. To prepare for future  survey cadence decisions and ensure that interesting and novel Solar System science is achievable with LSST, the SSSC has identified and prioritized key Solar System research areas for investigation with LSST in this roadmap. The ranked science priorities highlighted in this living document will inform LSST survey cadence decisions and aid in identifying  software tools and pipelines needed to be developed by the planetary community as added value products and resources before the planned start of LSST science operations. 
 \end{abstract}

\section{Introduction} 

Taking an inventory of the Solar System is one of the four key themes defining the science-driven requirements for the Large Synoptic Survey Telescope\footnote{LSST Science Requirements Document: \url{https://www.lsst.org/scientists/publications/science-requirements-document}} \citep[LSST;][]{2008arXiv0805.2366I}.  First light is expected in 2020, with full LSST science operations planned to commence in 2022. Objects brighter than approximately 16th magnitude will saturate in LSST observations, including all of the known Solar System's planets. Thus, the bulk of LSST's Solar System science will be derived from small body detections and observations. LSST will image and monitor millions of Solar System bodies. Over its 10-year lifespan, LSST is expected to catalog over 5 million Main Belt asteroids, almost 300,000 Jupiter Trojans, over 100,000 Near Earth Objects (NEOs), over 40,000 Kuiper belt objects  (KBOs), tens of interstellar objects, and over 10,000 comets \citep{2009arXiv0912.0201L,2010Icar..205..605S,2016ApJ...825...51C, 2017AJ....153..133E,2017ApJ...850L..38T}. Many of these objects will receive hundreds of observations in multiple bandpasses. LSST will report detections of moving objects in various filters (\emph{ugrizy}) between approximately 16 and 24.5 magnitudes (in r band) over its observing footprint (covering $\sim$18,000 square degrees in the Wide-Fast-Deep survey), link these detections into orbits, and provide metadata on observing conditions. 

It will be up to the planetary community to apply a wide variety of methods to synthesize and combine this information in order to fully leverage the LSST dataset for Solar System science. These goals include probing planetary formation and evolution and  placing the Solar System in context with other planetary systems. This requires an understanding of the current status of our Solar System -- the orbital and size distributions of small bodies (including asteroids, comets, planetary satellites, KBOs, and inner Oort cloud objects) and their physical properties (e.g., chemical composition, physical shape, mass, rotation rate, binarity, density, porosity, and mass loss rates). The LSST Solar System Science Collaboration (SSSC)\footnote{SSSC website: \url{http://www.lsstsssc.org}} aims to prepare methods and tools to analyze LSST data for Solar System science, as well as develop optimum survey strategies for discovering moving objects throughout the Solar System with LSST. 

We present a science roadmap that outlines the SSSC's ranked science priorities achievable with LSST during its planned baseline operations, expected to be 10 years. The list outlined here is not exhaustive but represents the most important Solar System-based research goals in the LSST era, based on the input from dedicated topical working groups. Crucial decisions about the LSST Wide-Fast-Deep survey cadence, special ancillary surveys that would maximize LSST science (mini-surveys), and target fields with deeper coverage and more frequent temporal sampling than the Wide-Fast-Deep survey (deep drilling fields) will be made over the next several years. This document serves as a guide for making these future cadence decisions. This roadmap will also help identify research areas where preparatory software tools and pipelines will need to be developed by the SSSC and broader community to produce data products beyond what the LSST project  will provide through  annual and nightly data releases. We expect this roadmap to be a living document that will be updated periodically as needed before LSST science operations commence.  

\section{What will LSST provide?}

The LSST project will deliver a Moving Object Processing System (MOPS) capable of identifying the bulk of transient moving Solar System bodies within LSST imaging data. An estimate of the predicted number of Solar System objects detected by LSST is detailed in the LSST Science Book\citep{2009arXiv0912.0201L}. The LSST Wide-Fast-Deep survey, with a proposed  Northern Ecliptic Spur mini-survey, would detect on the order of 100,000 NEOs, 5.5 million Main Belt asteroids, 280,000 Jupiter Trojans, and 40,000 KBOs with between 200-350 observations per object (for bright objects) in various filters. \citet{2010Icar..205..605S} estimate that LSST will discover approximately 10,000 comets, with 50 observations or more per object in various filters. \citet{2016ApJ...825...51C},  \cite{2017AJ....153..133E}, and \citet{2017ApJ...850L..38T} estimate that at least one interstellar object \citep[like `Oumuamua;][]{2017Natur.552..378M} is expected to be discovered by LSST annually. Additionally, detailed estimates of LSST NEO detection rates can be found in  \citet{2016AJ....151..172G}, \citet{2017AJ....154...13V}, and \citet{2017arXiv171110621J}. 

LSST's  basic capabilities will produce orbital distributions for large populations of Solar System objects, which can be de-biased using either the nominal completeness function or a user-supplied `truth' population (through the metadata on observing history), with precision photometry in multiple bandpasses and accurate astrometry. Beyond detection and orbit characterization, there will be multi-band photometry for each moving object detected, although in the wide-survey, objects will be measured in different filters at different times (with a variety of times between measurements that could vary from a few minutes to many days or months). There will be sparse light curves and photometric variability information, including an upper limit on an object's brightness when LSST does not detect a source, although these light curves will have variable time sampling. The catalogs will include measurements of the point-spread-function (PSF) of each source and a measurement of the deviation from the stellar PSF. There will also be tools to retrieve cutout images of each source to enable searches for outbursts, brightening events, and cometary activity and to perform trailed photometry (using a non-circular photometric aperture matched to the object's on-sky motion during the exposure) and forced photometry (performing photometry at a fixed/predicted position rather than fitting a centroid). Additionally, markers of comet-like activity or disruption events (including measures of source extendedness) will also be reported in a nightly public alert stream. 
 
\section{Overview of LSST Data Products and Data Access Services}

An overview of LSST  Data Management is provided in \cite{2015arXiv151207914J}. The summary below describes what LSST will provide for Solar System science. The LSST Data Products Definition Document\footnote{LSST Data Products Definition Document (DPDD):  \linebreak   \url{https://docushare.lsstcorp.org/docushare/dsweb/Get/LSE-163/LSE-163_DataProductsDefinitionDocumentDPDD.pdf}} (DPDD) is the authoritative source; here we provide a brief overview and interpretation of that document in the context of detecting and cataloging the small body populations residing within the Solar System.

LSST is expected to image the sky with a cadence\footnote{More about the observing footprint and plan can be found in the Operations Simulations documentation \cite[OpSim;][]{2014SPIE.9150E..15D} at \url{https://www.lsst.org/scientists/simulations/opsim} and simulations of the observing history can be found at \url{https://www.lsst.org/scientists/simulations/opsim/opsim-survey-data}.} generally appropriate for detecting small moving objects in the Solar System in \emph{ugrizy} filters over the Southern hemisphere. A proposed mini-survey would extend the LSST Wide-Fast-Deep survey coverage to the ecliptic plane + 10 degrees in the Northern hemisphere; this `Northern Ecliptic Spur' may only be imaged in \emph{griz}. The observing cadence is not fully set and may be revised based on feedback from the science collaborations and the broader community. We refer the reader to the LSST Observing Strategy White Paper \citep{2017arXiv170804058L} for more information.

The LSST project will create difference images where the LSST observations in each visit are subtracted from a template image to detect moving objects (plus transient and variable sources). Transient sources that are detected at 5-sigma or more above the sky background in the difference images will be identified and referred to as diaSources\footnote{Current LSST Database Schema for DiaSources: \url{https://lsst-web.ncsa.illinois.edu/schema/index.php?sVer=baseline&t=DiaSource}} in the LSST Database Schema, where DIA refers to Difference Image Analysis. Within 60 seconds of each observation, diaSources will be made public via a real-time stream of observation reports (alerts). These public alerts will include transients and variables and will not include linking between different visits, but known moving objects (as well as known variable sources) will be identified with coincident detections in the alerts. The alert information will include astrometry, photometry, and PSF shape information including trailing and direction of motion (to identify very fast moving Near-Earth objects [NEOs], even if unknown)  and identification of non-stellar PSFs (to identify outbursts or cometary activity).

Solar System bodies located at distances closer than $\sim$200 au will have sufficient on-sky motion between visits taken in a single night to be identified as moving objects by MOPS. MOPS identifies moving objects from the catalogs of diaSources. MOPS will link diaSources from each visit within one night into tracklets (potential linkages in the same night using linear extrapolation). Between nights, MOPS links tracklets into tracks (potential linkages over three nights using a quadratic fit). Only a track that can be fit to a heliocentric orbit with reasonable residuals is considered to be a reportable detection of a moving Solar System object (called an SSObject\footnote{Current LSST Database Schema for SSObjects: \url{https://lsst-web.ncsa.illinois.edu/schema/index.php?sVer=baseline&t=SSObject}} in the LSST database schema). All linked tracks and SSObjects will be reported daily to the Minor Planet Center\footnote{\url{https://www.minorplanetcenter.net}}. Additionally, MOPS will link additional diaSources with previously known or newly discovered SSObjects to extend the orbital arc for each moving object (as new images are taken or as new objects are discovered). 

The LSST project will provide access to the associated metadata stored with SSObjects as well as the observations themselves (called Sources\footnote{Current LSST Database Schema for Sources: \url{https://lsst-web.ncsa.illinois.edu/schema/index.php?sVer=baseline&t=Source}} when measured in images or diaSources when measured in difference images) in an online searchable database; this information will be available in yearly data releases as well as a daily database kept up to date with daily moving object processing. The yearly data releases will provide  measurements of each moving object detected with absolute astrometry accurate to better than 0.05$^{\prime\prime}$ and relative astrometry precise to 0.01$^{\prime\prime}$ over spatial scales of a few tens of arcminutes; absolute photometry will be accurate to 10 mmag for bright (\emph{r}$<$20) sources and relative photometry precise to 5 mmag for observations over spatial scales small relative to the visible sky in griz (7 mmag in \emph{uy}). 

Through a Science User Interface\footnote{LSST Science User Interface $\&$ Science User Tools Conceptual Design: \url{https://docushare.lsst.org/docushare/dsweb/Get/Rendition-22538}} (SUI), users will have access to LSST catalog and image query tools. Users will be able to access metadata on observing conditions, such as the telescope pointing history, seeing history, cloud conditions, and estimated 5-sigma limiting magnitudes in the difference images for each visit. Through the SUI, users will be able to access catalog entries or postage stamps associated with each diaSource and/or Source for a particular SSObject as requested, as well as receive postage stamps from LSST images which match a user-defined orbit or user-defined RA (right ascension)/Dec(declination)/time regions.  Some computing resources through  data centers will be available for extended analysis of catalogs or images where the analysis routines are written by users and interface with application programming interfaces (APIs) provided by the LSST project.

\section{Science Priorities of the SSSC}
\label{sec:roadmap}
In the following section, we briefly outline the ranked LSST science priorities as determined by the SSSC membership. We divide the small body populations of the Solar System in four broad categories based on location and similar discovery challenges:
\begin{itemize}
\item{Active Objects - broadly consisting of all categories of activity in the small body populations  (i.e., objects exhibiting some type of mass loss): short period comets, long period comets, Main Belt comets, impact-disrupted or rotationally-generated active asteroids, etc.}
\item{Near Earth Objects (NEOs) and Interstellar Objects- broadly consisting of objects on orbits inward of or diffusing inward from the asteroid belt and objects on unbound orbits passing through the Solar System, like `Oumuamua \citep{2017Natur.552..378M}.} 
\item{Inner Solar System - broadly consisting of Main Belt asteroids, Hildas, Mars/Jupiter Trojans, and Mars/Jupiter satellites.}
\item{Outer Solar System - broadly consisting of Kuiper belt objects (KBOs), Centaurs, inner Oort cloud (Sedna-like objects), Oort cloud objects, Saturn/Uranus/Neptune Trojans, and Saturn/Uranus/Neptune satellites.}
 \end{itemize}
For each of these core small body populations, the SSSC has a list of science goals to achieve with LSST, ranked from highest priority to lowest. 

\subsection{Active Objects}
\begin{enumerate}

\item{Discovery and orbital classification of large numbers of active objects to understand and model the onset and termination of activity in the different Solar System small body populations and to explore correlations between physical/orbital characteristics and transient activity.}

\item{Discovery/frequency/population estimates of coma and/or dust tail-bearing bodies in the Solar System small body populations  including  Main-belt comets, NEOs, collisionally impacted asteroids, Centaurs, KBOs, short and long period  comets, and interstellar objects to better probe the drivers of such activity and measure the size of these reservoirs.}

\item{Detection/frequency/population estimates of anomalous outbursts and rapid brightening/splitting events above the expected brightness evolution of objects in the Solar System small body populations to better probe the drivers of such activity and the size of these reservoirs.}

\item{Characterization of the changes in color, morphology, brightness, rotation, shape, and other observable properties of active objects over time (including changes from pre-activity/outburst properties) and at different epochs in the orbits of these bodies to probe surface changes and better explore the various drivers of such activity and their evolution.}

\item{Determination of rotational light curves for a large sample of active objects to study physical properties of active objects, including the spin angular momentum distribution, shape distribution, and binary frequency.}

\item{Detection and characterization of the non-gravitational forces  (including jet-driven and collisional accelerations) acting on active bodies to compute better original and future orbits (especially important for identifying dynamically new or long period comets) and  estimate rotation poles and  seasonal states of active body nuclei.}

\end{enumerate}

\subsection{Near-Earth Objects (NEOs) and Interstellar Objects}
\begin{enumerate}

\item{Compilation of an NEO catalog with high completeness and adequate orbit quality.}

\item{Color measurements and broad phase coverage of NEOs, including distinguishing NEOs of cometary origin through color measurements and  probing the color distribution of ten-meter scale objects.}

\item{Timely advance notice of close approaches or potential impacts to facilitate time critical characterization efforts including radar, spectroscopic, and light curve observations.}

\item{Measurement of the orbital, absolute magnitude, and taxonomy distributions within the NEO population, enabling the identification of correlations between taxonomy and orbital properties for all NEOs and the determination of the orbital distribution of ten-meter scale objects.}

\item{Determination of the long-term impact flux of NEOs as a function of size, for $\ge$ 140 m bodies in particular.}

\item{Discovery/frequency/population estimates of interstellar objects on unbound orbits passing through the Solar System as a potential probe of planet formation and planetesimal ejection rates in the local solar neighborhood.}

\item{Determination of rotational light curves for a large sample of NEOs to study physical properties of NEOs, including the spin angular momentum distribution, shape distribution, and binary frequency.}

\item{Detection and characterization of the non-gravitational forces (including the Yarkovsky effect, solar radiation pressure, outgassing, collisions) acting on NEOs to explore and better understand how NEO orbits evolve over time.}

\item{Measurement of the absolute magnitude distribution of temporarily-captured objects (NEOs that are  temporarily captured by the gravity-well formed by the Earth and Moon) in order to compare to model predictions and to probe the low end of the asteroid size/absolute magnitude distribution.}

\item{Investigation of the possible NEO disruption mechanisms active at small perihelion distances to probe NEO internal structure and test dynamical models.}

\end{enumerate}

\subsection{Inner Solar System}
\begin{enumerate}

\item{Discovery and orbital classification of large numbers of asteroids and  Mars/Jupiter Trojans to probe their orbital and absolute magnitude distributions and to measure the size frequency distributions of different taxonomic classes.}
\item{Measurement of high quality astrometry for new and previously known asteroids, Mars/Jupiter Trojans, and Jupiter irregular satellites to refine orbits and improve ephemerides for stellar occultation predictions.}
\item{Detection of impacts of small asteroids onto large ones and detection of asteroid disruption by impact to probe the current collisional environment within the asteroid belt, study dust dynamics, constrain asteroid  internal structure,  and explore space weathering processes through comparison of surfaces before and after detected impacts.}
\item{Determination of colors and compositions for a large sample of asteroids, specifically including Jupiter's irregular satellites, Mars/Jupiter Trojans, Hildas, Cybeles, and Phobos and Deimos to identify correlations with dynamical  and taxonomic information with implications for understanding the formation of the inner solar system (e.g., chemical distribution in the primordial disk; collisional family parent bodies and formation events). }
\item{Investigation of the hydration of C-complex objects and Main Belt asteroids to explore the compositional evolution of the inner solar system and test giant planet migration models.}
\item{Determination of rotational light curves for a large sample of asteroids in different taxonomic classes to study physical properties of asteroids,  including the spin angular momentum distribution, shape distribution, and binary frequency.}
\item{Improved characterization of newly discovered and previously known asteroid families, clusters, and pairs to study genetic relationships and homogeneity of collisional families at small sizes.}
\item{Measurement of asteroid masses and bulk densities from mutual gravitational interactions to probe asteroid internal structures and test  planet formation models.}
\item{Detection and frequency of rotational fission within the non-NEO asteroid populations to probe internal structure and test dynamical models.}
\end{enumerate}

\subsection{Outer Solar System}
\begin{enumerate}

\item{Discovery and orbital classification of large numbers of outer Solar System objects over a wide range of sizes (H$>$9) and orbits to characterize the size-frequency-orbit distribution of KBOs and to probe the formation and evolution of the outer solar system (e.g., comet/Centaur pathways, collisional evolution, Neptune migration, etc.).}

\item{Discovery and orbital classification of objects on unusual or extreme orbits, especially inner Oort cloud objects (i.e. Sedna-like objects) with high perihelia ($q$ $>$ 40 au) and objects with very high inclination ($i$ $>$ 40 deg), to place constraints on proposed origin scenarios   \citep[e.g., the putative Planet 9;][]{2014Natur.507..471T,2016AJ....151...22B,2016ApJ...824L..23B,2016AJ....152..221S,2017AJ....154...65B}. }

\item{Determination of colors for large numbers of objects to identify correlations with dynamical information with implications for understanding the formation of the outer solar system (e.g., chemical distribution in the primordial disk; collisional families). }

\item{ Determination of rotational light curves for large numbers of objects from different dynamical classes to study physical properties of KBOs, including spin angular momentum distribution and binary frequency. }

\item{Discovery and orbital classification of large numbers of objects in resonance with the giant planets, especially the libration islands of high-order resonances of Neptune, to constrain models of Neptune migration. }

\item{Discovery and clear characterization (e.g., PSF shape) of binaries and multiple systems wide enough to be resolved or partially resolved. }

\item{Measurement of acccurate and precise astrometry for known and new distant Solar System bodies to enable stellar occultation observations. }

\end{enumerate}

\section{Community Software Development}

Given the unprecedented scale of the LSST survey, tools that can conduct rapid automated analyses of the large quantities of data produced on a nightly basis will be essential for taking full advantage of LSST's scientific potential. Furthermore, a collaborative approach will almost certainly be required within the planetary community to ensure that the broad range of analysis tools and software pipelines that will be required are available and ready to be implemented by the time  LSST begins full science operations.  In this Section, we briefly describe the status of the SSSC's community software and infrastructure development. 

One of the key objectives of the software development process being undertaken by the SSSC is the identification of common software and analysis needs among the planetary community based on the list of scientific priorities detailed in Section \ref{sec:roadmap} in order to organize a coordinated effort that best maximizes available resources and effort.  In some cases, this development will involve the adaptation or automation of existing software pipelines, while in other cases, entirely new tools will need to be developed.  LSST will provide some computational resources  for the analysis of LSST images and associated data products with user-added software. To make the most of these limited computational resources, it will be in the planetary community's interest to aggregate the various  community developed tools that require access to LSST image data and apply them to each image of a moving object detection at the same time. This will require  image data to be retrieved only once, minimizing the computational draw on the LSST servers.  A key goal of the SSSC's software development effort is to create an overarching  database that will store all user-derived values and output associated with LSST moving object discoveries produced by SSSC analysis tools. To facilitate SSSC investigations that take full advantage of the diverse higher-level data products that are expected to be generated from LSST observations, we are planning for this database to be fully accessible through APIs and a web-based query form.

\section{Future Work}
This science roadmap serves as a starting point towards maximizing LSST's  potential for Solar System science. The extent to which the above science priorities will be achieved is directly related to the final survey strategy and cadence selected. The next step for the SSSC will be to generate quantifiable success metrics for each of the above science priorities that can be tested against the various LSST Wide-Fast-Deep survey observing strategies and their associated simulated observing histories. Additionally, one of the next steps for the SSSC will be to identify which of the above science goals would be significantly enhanced by or would be only achievable through a specially designed mini-survey (such as the proposed `Northern Ecliptic Spur') or deep drilling field.  Further actions include identifying and beginning development on user-added community software tools and pipelines needed beyond what the LSST project will provide in order to carry out the desired science goals in this roadmap. 

\section*{Acknowledgements}

The authors thank the  Large Synoptic Survey Telescope (LSST)  Project Science Team and the LSST Corporation for their support of LSST Solar System Science Collaboration's (SSSC) efforts. This work was supported in part by a LSST Corporation Enabling Science grant.  MES was supported by Gemini Observatory, which is operated by the Association of Universities for Research in Astronomy, Inc., under a cooperative agreement with the National Science Foundation  on behalf of the Gemini partnership: the National Science Foundation (United States), the National Research Council (Canada), CONICYT (Chile), Ministerio de Ciencia, Tecnolog\'{i}a e Innovaci\'{o}n Productiva (Argentina), and Minist\'{e}rio da Ci\^{e}ncia, Tecnologia e Inova\c{c}\~{a}o (Brazil). This manuscript was prepared using the AASTex latex classfile and template package from America Astronomical Society (AAS) Journals \url{http://journals.aas.org/authors/aastex/aasguide.html}. This work has made use of NASA's Astrophysics Data System Bibliographic Services. The authors thank Amy Barr and Bryce Bolin for constructive comments on the manuscript.

\bibliographystyle{aasjournal}
\bibliography{ms} 

\end{document}